\documentclass{aastex63}
\shorttitle{Mutual Inclination of USPs}
\shortauthors{Chen et al.}

\usepackage{amsmath}
\usepackage{placeins}
\usepackage{xcolor}
\usepackage{soul}

\begin{document}

\title{Mutual Inclination of Ultra-Short-Period Planets with Time Varying Stellar J2-moment}

\correspondingauthor{Chen Chen}
\email{chen\_chen@gatech.edu}

\author[0000-0002-0231-5124]{Chen Chen}
\affiliation{School of Earth and Atmospheric Sciences,
Georgia Institute of Technology,
Atlanta, GA, 30332, USA}

\author{Gongjie Li}
\affiliation{School of Earth and Atmospheric Sciences,
Georgia Institute of Technology,
Atlanta, GA, 30332, USA}
\affiliation{Center for Relativistic Astrophysics,
School of Physics,
Georgia Institute of Technology,
Atlanta, GA 30332, USA}

\author[0000-0003-0412-9314]{Cristobal Petrovich}
\affiliation{Instituto de Astrofísica, Pontificia Universidad Católica de Chile, Av. Vicuña Mackenna 4860, 782-0436 Macul, Santiago, Chile}
\affiliation{Millennium Institute for Astrophysics, Chile}





\begin{abstract}

Systems with ultra-short-period planets (USPs) tend to possess larger mutual inclinations compared to those with planets located farther from their host stars. This could be explained due to precession caused by stellar oblateness at early times when the host star was rapidly spinning. However, stellar oblateness reduces over time due to the decrease in the stellar rotation rate, and this may further shape the planetary mutual inclinations. In this work, we investigate in detail how the final mutual inclination varies under the effect of a decreasing $J_2$. We find that different initial parameters (e.g., the magnitude of $J_2$ and planetary inclinations) will contribute to different final mutual inclinations, providing a constraint on the formation mechanisms of USPs.
In general, if the inner planets start in the same plane as the stellar equator (or co-planar while misaligned with the stellar spin-axis), the mutual inclination decreases (or increases then decreases) over time due to the decay of the $J_2$ moment. This is because the inner orbit typically possesses less orbital angular momentum than the outer ones. However, if the outer planet is initially aligned with the stellar spin while the inner one is misaligned, the mutual inclination nearly stays the same. Overall, our results suggest that either the USP planets formed early and acquired significant inclinations (e.g., $\gtrsim30^\circ$ with its companion or $\gtrsim10^\circ$ with its host star spin-axis for Kepler-653c) or they formed late ($\gtrsim$Gyr) when their host stars rotate slower.

\end{abstract}



\section{Introduction} \label{sec:intro}

The ultra-short-period planets (USPs) refer to planets orbiting their host stars with periods shorter than one day. They typically have radius less than $2 R_{\oplus}$ and orbit around $\sim 0.5\%$ of G-dwarf stars as well as $\sim 0.8\%$ K-dwarf stars \citep{Sanchis_Ojeda_2014, winn_sanchis-ojeda_rappaport_2018}. As reported by \cite{Dai_2018}, planets with smaller orbital distances ($a/R_{\star} < 5$) have higher mutual inclinations with exterior planets than those with larger orbital distances ($5 < a/R_{\star} < 12$). They also found that higher mutual inclinations between the planets are correlated with larger period ratios, showing that USP systems are typically hierarchical.

As the orbital distances of the USP planets are within the dust sublimation zone \citep{flock2019}, the formation of these extreme objects remains puzzling. Previous works have proposed that the USP planet initially forms on a wider orbital distance and then migrates inward by some mechanisms. For instance, the planet can form in-situ and the orbital distance is shrunk by the tidal effects \citep{Lee_2017}. The dynamical processes, which involve either high eccentricity \citep{Petrovich_2019} or low eccentricity migration \citep{Pu_2019}, can also contribute to the inward migration and large mutual inclination of USP planets. Planet obliquity tides can also produce USP planets with low initial semi-major axes ($a\lesssim0.05$ au) \citep[][]{Millholland_Spalding_2020}. Moreover, during episodic accretion events, the planet migrates into a USP orbit in a very short timescale by headwind torques \citep{becker2021migrating}.

The oblateness of the central star plays an important role in the dynamics of planetary systems with small orbital distances, especially for the USPs. Recently, \cite{Li_2020} showed that the stellar oblateness can explain the mutual inclination of the USPs identified by \cite{Dai_2018}, assuming the initial configuration is co-planar. Considering the stellar quadrupole moment and a planetary companion as two mechanisms, \cite{Becker_2020} suggested that these two processes can produce a misalignment between the USP and the tightly packed co-planar planets. Moreover, \cite{Spalding_2020} found that the stellar oblateness has a stronger influence than a distant giant and it is able to excite the mutual inclination with a rapid disk-dispersal, and \cite{Schultz21} found that a large $J_2$ could enhance mutual inclination and lead to orbital instability. For general planetary systems, \cite{Spalding_2016} found that the significant misalignment between the orbital planes can be excited and the system can also undergo the dynamical instability for planets orbiting around a tilted star with a decreasing stellar oblateness ($J_2$). This helps to explain the $Kepler$ Dichotomy.

As the stellar oblateness plays an important role in the dynamics of the USPs, we investigate the evolution of inclinations of the USPs system due to the stellar $J_2$ in more detail. We show that observed mutual inclination between planets could help constrain the formation mechanisms of the USPs.
In particular, we investigate the contribution due to a decaying stellar oblateness, as the rotation rate of the star reduces by magnetic braking. For simplicity, we only consider the effects of stellar oblateness and planet-planet interaction, assuming the disk has been largely dissipated before the arrival of the USPs. The mutual inclination changes with different initial conditions, and thus the observed minimum mutual inclination could be used to constrain the formation channels of the USPs \citep[as suggested by][]{Becker_2020} as well as the initial configuration of the system. We use Kepler-653, a two-planet system with a large mutual inclination, as an illustrative example in this paper. 

We organize the paper as the following: in Section \ref{sec:method}, we investigate the evolution of mutual inclination for general systems with a USP using the secular approach. In Section \ref{sec:j2_section}, we apply our results to the specific system Kepler-653 to constrain its formation channel. In Section \ref{sec:disc_conclu}, we summarize and discuss the possible formation scenarios for Kepler-653.

\section{Analytical and simulation results} \label{sec:method}

In this article, we consider the dynamics of a USP planet orbiting an oblate star while being perturbed by a farther planetary companion, and we assume the disk has been dissipated for simplicity. We study how the mutual inclination of the two planets evolves as the stellar $J_2$ moment decays. We detail the set up of our problem and present the Hamiltonian in Section \ref{sec:secular_hami}, then we use the secular regime to analyze the dynamics in Sections \ref{sec:hami} and \ref{sec:beta}. 

\subsection{Secular Hamiltonian} \label{sec:secular_hami}

We consider a planetary system with two planets and assume the orientation of stellar spin is unchanged. The fixed stellar spin orientation is a good approximation for a relatively compact and low mass system, since the angular momentum of the star dominates, as illustrated by \cite{Spalding_2016}.
For a 3-body system with a USP and an outer planet, the ratio of semi-major axis tends to be large (i.e., the period ratio $P_2/P_1 \gtrsim 5$), as suggested by \cite{Dai_2018}. So we consider the doubly time-averaged planet-planet interaction potential up to octupole order in dimensionless form, which is well documented in the literature \citep[e.g.,][]{Fabrycky07, naoz13, Petrovich_2015} 
:

\begin{equation}
\begin{aligned}\label{eq:planet_potential}
\phi_{\rm planet} = \frac{\phi_0}{(1-e_o^2)^{3/2}}[\frac{1}{2}(\mathbf{j}_i\cdot\hat{\mathbf{j}}_o)^2 + (e_i^2-\frac{1}{6})-\frac{5}{2} (\mathbf{e}_i\cdot \hat{\mathbf{j}}_o)^2] \\
\quad + \frac{25 a_i \phi_0}{16 a_o (1-e_o^2)^{5/2}} \{ (\mathbf{e}_i\cdot\mathbf{e}_o) [(\frac{1}{5} - \frac{8}{5}e_i^2) - (\mathbf{j}_i\cdot\hat{\mathbf{j}}_o)^2 \\
+ 7(\mathbf{e}_i\cdot\hat{\mathbf{j}}_o)^2] - 2(\mathbf{j}_i\cdot\hat{\mathbf{j}}_o) (\mathbf{e}_i\cdot\mathbf{\hat{j}}_o)(\mathbf{j}_i\cdot\mathbf{e}_o) \},
\end{aligned}
\end{equation}
where the subscriptions $i$ and $o$ represent inner and outer planet, $\mathbf{e}_i$ and $\mathbf{e}_o$ are the eccentricity vectors, $ \mathbf{j}_i = (1-e_i^2)^{1/2} \hat{\mathbf{j}}_i$ and $ \mathbf{j}_o = (1-e_o^2)^{1/2} \hat{\mathbf{j}}_o$ are the dimensionless orbital angular momentum vectors with unit vectors $\mathbf{\hat{j}}_i$, $\mathbf{\hat{j}}_o$, and

\begin{equation}
    \phi_0 = \frac{3G m_i m_o a_i^2}{4a_o^3}
\end{equation}

In addition, due to the rotational deformation, the oblate star contributes a quadrupole potential, which can be expressed as the following, \citep[e.g.,][]{tremaine_yavetz_2014}

\begin{equation}\label{eq:j2_potential}
    \phi_{J2} = \frac{G M_s J_2 R_s^2}{4 a^3 (1-e^2)^{3/2}}[1-3(\hat{n}_s \cdot \mathbf{\hat{j}})^2]
\end{equation}
where $M_S$ is the stellar mass, $R_s$ is the radius of the star, $\hat{n}_s$ is the unit vector of the stellar spin axis, and $\hat{\mathbf{j}}$ is the unit vector of the orbital angular momentum. Assuming $e = 0$ in Equation \ref{eq:j2_potential}, the $J_2$ precession frequency scaled by the mean motion can be written as

\begin{equation}\label{eq:j2_freq}
    \frac{\dot{\Omega}}{n} = -\frac{3}{2} J_2 \left(\frac{R_s}{a}\right)^2 \cos I,
\end{equation}
where $I$ is the inclination between planetary orbital angular momentum and the stellar spin axis. It is clear from Equation \ref{eq:j2_freq} that $J_2$ precession plays an important role in the dynamics of innermost planet as $|\dot{\Omega}| \varpropto a^{-3.5}$. Combining Equations \ref{eq:planet_potential} and \ref{eq:j2_potential}, we obtain the secular Hamiltonian:

\begin{equation}
\begin{aligned}\label{eq:full_H1}
    \mathcal{H} = \frac{\phi_0}{(1-e_o^2)^{3/2}}[\frac{1}{2}(\mathbf{j}_i\cdot\hat{\mathbf{j}}_o)^2 + (e_i^2-\frac{1}{6})-\frac{5}{2} (\mathbf{e}_i\cdot \hat{\mathbf{j}}_o)^2] + \frac{25 a_i \phi_0}{16 a_o (1-e_o^2)^{5/2}} \{ (\mathbf{e}_i\cdot\mathbf{e}_o) [(\frac{1}{5} - \frac{8}{5}e_i^2) - (\mathbf{j}_i\cdot\hat{\mathbf{j}}_o)^2 \\
\quad + 7(\mathbf{e}_i\cdot\hat{\mathbf{j}}_o)^2] - 2(\mathbf{j}_i\cdot\hat{\mathbf{j}}_o) (\mathbf{e}_i\cdot\mathbf{\hat{j}}_o)(\mathbf{j}_i\cdot\mathbf{e}_o) \} 
-\frac{G M_s m_i J_2 R_s^2}{4 a_i^3 (1-e_i^2)^{3/2}}[1-3(\hat{n}_s \cdot \mathbf{\hat{j}}_i)^2]
-\frac{G M_s m_o J_2 R_s^2}{4 a_o^3 (1-e_o^2)^{3/2}}[1-3(\hat{n}_s \cdot \mathbf{\hat{j}}_o)^2],
\end{aligned}
\end{equation}

USP orbits are typically circular due to fast tidal circularization timescales \citep{winn_sanchis-ojeda_rappaport_2018}, and the timescales of orbital decay and spin-alignment on the inner planet are much longer than those of planet-planet interaction (e.g., \cite{Rodriguez_2018, Becker_2020}), thus we assume the planets to be near circular and we neglect tidal effects in this work. We note that larger mutual inclination above $\sim 40^\circ$ could lead to eccentricity excitation of the USPs due to von Zeipel-Lidov-Kozai oscillations, and this could also lead to instability \citep[e.g.,][]{Spalding_2016, Schultz21}, when the General Relativity (GR) effect is not sufficient to prevent the eccentricity excitation \citep[e.g.,][]{faridani2021hiding}. Thus, we only consider lower mutual inclinations below $40^\circ$ with near circular orbits in our study.
The first order post-Newtonian (1PN) correction for GR effects is not included in our work, because it only causes the argument of pericenter to precess without affecting the mutual inclination \citep[e.g.,][]{Li_2020}.

\subsection{Energy Contours with Different J2 Values}\label{sec:hami}
In this section, we analyze the secular results using the contours of constant Hamiltonian. For simplicity, we assume the orbits are circular (i.e., $e_i = e_o = 0$) throughout the evolution. We relax this assumption in Section \ref{sec:j2_section}. Based on Equation \ref{eq:full_H1}, the Hamiltonian of circular case can be represented by
\begin{equation}\label{eq:H_c}
    \mathcal{H}_{\rm circular} = \frac{\phi_0}{2} (\mathbf{\hat{j}}_i \cdot \mathbf{\hat{j}}_o)^2 
    -\frac{G M_s m_i J_2 R_s^2}{4 a_i^3}[1-3(\hat{n}_s \cdot \mathbf{\hat{j}}_i)^2] 
    -\frac{G M_s m_o J_2 R_s^2}{4 a_o^3}[1-3(\hat{n}_s \cdot \mathbf{\hat{j}}_o)^2].
\end{equation}

The dynamics can vary with the ratio of the angular momenta of the inner and the outer orbits. For instance, \cite{Spalding_2016} showed that if the inner planet has more orbital angular momentum than the outer one, secular resonance could occur and lead to larger mutual inclination. The ratio of the angular momentum between inner and outer planet can be represent by \citep[e.g.,][]{Petrovich_2018},
\begin{equation}\label{eq:beta}
    \beta = \frac{m_i a_i^{1/2}}{m_o a_o^{1/2}}.
\end{equation}
To illustrate how the different angular momentum ratios change the dynamics, we adopt two systems with different ratios ($\beta = 0.05$ for System A and $\beta = 5$ for System B) in the following. The parameters for these systems are shown in Table \ref{tab:hypo}. We use the solar mass and radius for both systems. 

During the evolution of the system, the z-component of angular momentum $\mathcal{J}_z$ (along the direction of the stellar spin-axis) is conserved under the effect of $J_2$. Normalizing $\mathcal{J}_z$ with respect to the outer orbit angular momentum, $\overline{\mathcal{J}_z}$ can be written as \citep{Petrovich_2018},  
\begin{equation}\label{eq:angular_momentum}
    \overline{\mathcal{J}_z} = \beta \cos I_i + \cos I_o = const.
\end{equation}
Applying the conservation of the z-component of angular momentum, the system can be reduced to one degree of freedom\footnote{ As shown by \citet{petro2020}, for small inclinations the Hamiltonian reduces to the second model for the resonance \citep{HL1983}}. Thus the evolution of the system can be described in a two dimensional space in terms of the orbital inclination and the differences in the longitude of ascending node. We then plot the constant energy contours to illustrate the dynamics over a large parameter space. 

\begin{deluxetable*}{ccc}
\tablecaption{Parameters of Hypothetical USP systems \label{tab:hypo}}
\tablewidth{0pt}
\tablehead{
\colhead{} & \colhead{\textbf{System A}} & \colhead{\textbf{System B}}
}
\startdata
Mass $m_i$ (\textit{$M_{\rm Earth}$}) & 1 & 10\\
Mass $m_o$ (\textit{$M_{\rm Earth}$}) & 10 & 1\\
Semimajor axis \textit{$a_i$} (AU) & 0.01  & 0.01\\
Semimajor axis \textit{$a_o$} (AU) & 0.04  & 0.04\\
Eccentricity \textit{$e_i = e_o$} & 0 & 0\\
$\beta$ & $ 0.05$ & $ 5$ \\
\enddata
\end{deluxetable*}

\begin{figure*}[ht!]
\centerline{\includegraphics[height=3.6 in]{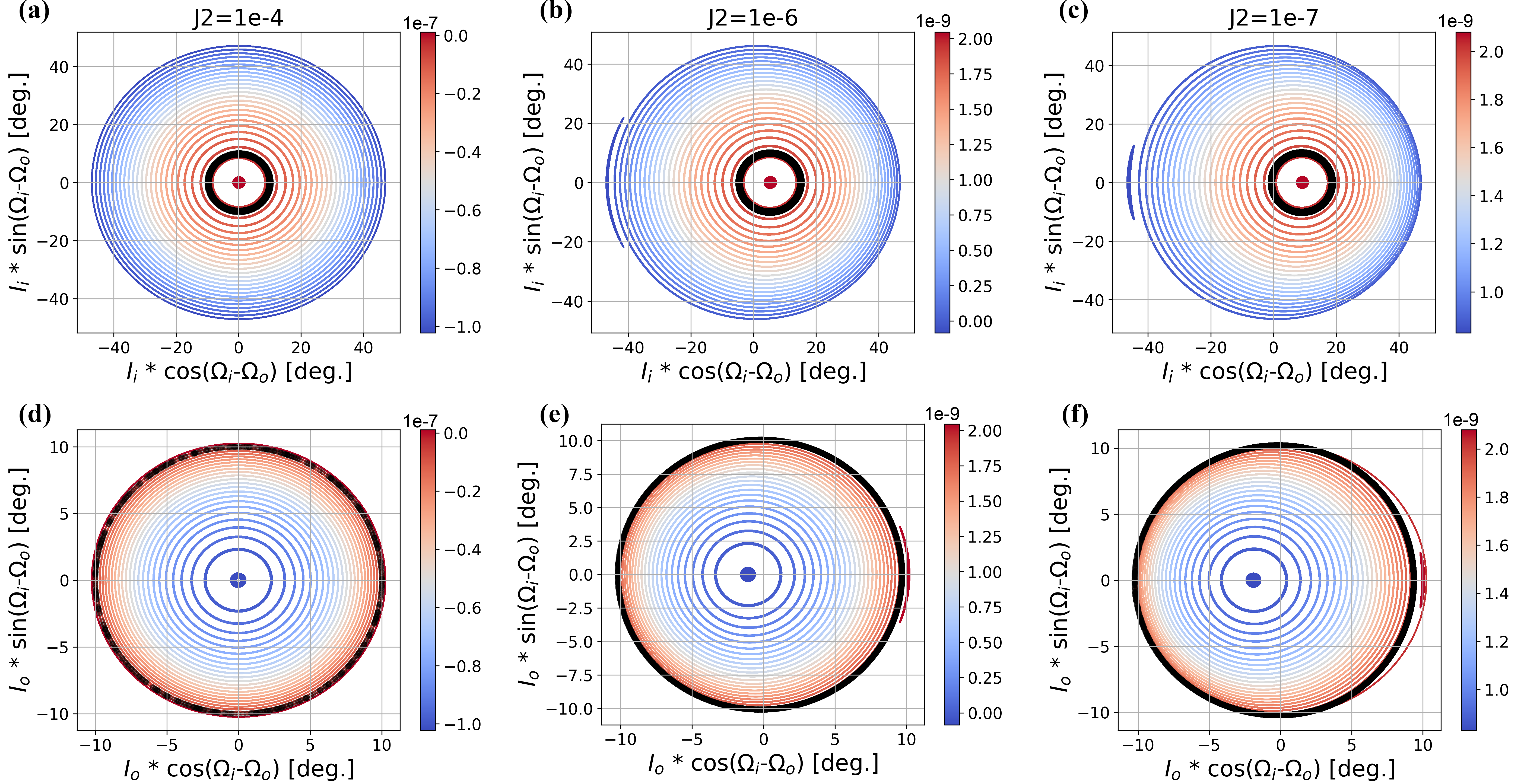}}
\caption{Energy contours of constant Hamiltonian for System A (Equation \ref{eq:H_c}) with different J2 value.  All the contours correspond to the same $\overline{\mathcal{J}_z}$. 
 Different columns correspond to different $J_2$ values. The x axis is $I_{\rm inner} \cos(\Omega_{\rm inner}-\Omega_{\rm outer})$ and the y axis is $I_{\rm inner} \sin(\Omega_{\rm inner}-\Omega_{\rm outer})$ in the first row for the inner planet. For the second row, the x axis is $I_o \cos(\Omega_i-\Omega_o)$ and the y axis is $I_o \sin(\Omega_i-\Omega_o)$ for the outer planet. The color represent the total potential energy $\mathcal{H}_{\rm circular}$ and the black dots show the trajectory of a secular evolution starting with the coplanar configuration ($I_{\rm inner,0} = 10^{\circ}$ and $I_{\rm outer,0} =10^{\circ}$). \label{fig:contours}}
\end{figure*}

\subsubsection{System dominated by the outer angular momentum}\label{sec:sys_a}

First, we show System A ($\beta = 0.05$), where the outer planet possesses more orbital angular momentum. We set $\overline{\mathcal{J}_z}$ to be the same for all the runs that we include in the Figure \ref{fig:contours}, and $\overline{\mathcal{J}_z}$ is calculated using Equation \ref{eq:angular_momentum} for the case with $I_{\rm inner}=I_{\rm outer}=10^{\circ}$ and $\beta = 0.05$. Different columns of Figure \ref{fig:contours} represent different $J_2$ levels. The first row of Figure \ref{fig:contours} shows the contours for inner planet inclination, which is in the plane of $[I_i \cos(\Omega_i-\Omega_o)$, $I_i \sin{(\Omega_i-\Omega_o)}]$, while the second row shows the contours for outer planet inclination, which is in the plane of $[I_o \cos(\Omega_i-\Omega_o)$, $I_o \sin{(\Omega_i-\Omega_o)}]$. Note that the axes correspond to the Cartesian Poincaré coordinates when the inclinations are low. The color represents the value of $\mathcal{H}_{\rm circular}$. The black dots come from a numerical solution (to be discussed later in this section), and they represent how the trajectory of the planets evolves as J2 decays over time if the planets start coplanar.
The energy contours are determined by the value of $\overline{\mathcal{J}_z}$ and $\mathcal{H}_{circular}$. We note that the parameter space with the mutual inclinations larger than $\sim 40^{\circ}$ might not be valid, as the Kozai cycle could excite the eccentricities with the large mutual inclination but we fix the eccentricities to be zero in this case. 

We can see from Figure \ref{fig:contours} that the aligned fixed point ($|\Omega_i-\Omega_o| = 0^{\circ}$ marked as the red dot) of the inner planet gradually moves from the center (i.e., star-aligned) to the right (i.e., planet-aligned, $\sim 9^{\circ}$) as $J_2$ decreases. The anti-aligned fixed point ($|\Omega_i-\Omega_o| = 180^{\circ}$ marked as the blue dot) of the outer planet increases slightly in inclination. It can also be observed that if the system is initially further from the fixed point, the oscillation amplitude of the inclination is larger.

The fixed points with the aligned nodes correspond to the case of an aligned inner planet (with respect to the stellar spin-axis) when the initial $J_2$ is large   , and those with the anti-aligned nodes correspond to the case with an aligned outer planet. The effect of $J_2$ on these fixed points shows us the inclination evolution of the planetary systems metioned above. To illustrate the evolution of fixed point with a decaying $J_2$, we adopt a similar approach as \cite{Petrovich_2018}, while similar analysis have been made earlier in the context of Cassini state \citep{BOUE2006312,refId0, anderson_lai_2018}. Ignoring the $J_2$ potential of the outer planet and considering the equilibrium condition of $\frac{d(\Omega_{\rm inner} - \Omega_{\rm outer})}{dt}=0$, the evolution of aligned fixed point ($\Omega_{\rm inner}-\Omega_{\rm outer} = 0$) can be found by
\begin{equation}\label{eq:fix_point_eq_aligned}
    \frac{2 a^5_L}{a_{\rm inner}^5 \beta} \sin I_{\rm outer} \sin 2I_{\rm inner} = -\Big(\frac{\sin I_{\rm outer}}{\beta} + \sin I_{\rm inner}\Big)\sin[2(I_{\rm inner}-I_{\rm outer})] 
\end{equation}
The evolution of anti-aligned fixed point ( $|\Omega_c-\Omega_b| = 180^{\circ}$) is represented by,
\begin{equation}\label{eq:fix_point_eq_mis}
    \frac{2 a^5_L}{a_{\rm inner}^5 \beta}  \sin I_{\rm outer} \sin 2I_{\rm inner} = \Big(-\frac{\sin I_{\rm outer}}{\beta} + \sin I_{\rm inner}\Big)\sin[2(I_{\rm inner}+I_{\rm outer})] 
\end{equation}

where the Laplace radius $a_L$ \citep[e.g.,][]{Tremaine_2009, Tamayo13} is 
\begin{equation}
    a^5_L = \frac{J_2 a^3_{\rm outer} M_s R^2_S}{m_{\rm outer}}
\end{equation}

\begin{figure}[ht!]
\centerline{\includegraphics[height=1.7 in]{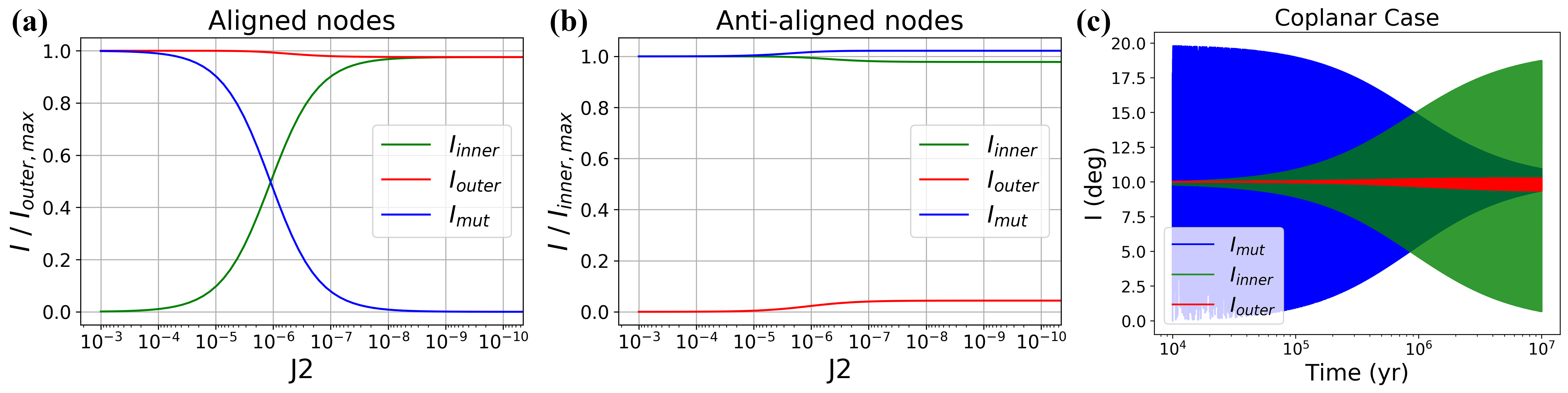}}
\caption{ (a) Evolution of the aligned fixed point for the inclinations based on the Equation \ref{eq:fix_point_eq_aligned} with respect to the decreasing $J_2$. The y-axis is the ratio between the inclination and the maximum outer inclination (i.e., when $J_2$ is $\sim 10^{-3}$). (b) Evolution of the anti-aligned fixed point for the inclinations based on the Equation \ref{eq:fix_point_eq_mis} with respect to the decreasing J2. The y-axis is the ratio between the inclination and the maximum inner inclination (i.e., when $J_2$ is $\sim 10^{-3}$). (c) Secular evolution for an initially co-planar case with $I_{\rm inner,0}=I_{\rm outer,0}=10^{\circ}$. \label{fig:fix_point}}
\end{figure}

Then, combining with Equation \ref{eq:angular_momentum} the fixed points can be determined at any given stellar $J_2$. The evolution of the aligned fixed point is shown in Figure \ref{fig:fix_point} (a). The maximum outer planet inclination is used as reference,
\begin{equation}\label{eq:fix_point_eq}
    I_{\rm outer, max} = \cos^{-1} \left(\overline{\mathcal{J}_z} - \beta\right)
\end{equation}
i.e., we plot the aligned fixed point inclination normalized by the maximum outer planet inclination, which corresponds to the initial outer inclination (initial inclination here refers to the fixed point inclination when $J_2$ is large, e.g., $J_2 \sim 10^{-3}$). Figure \ref{fig:fix_point} (a) shows that the fixed point of the inner inclination increase from zero to $\sim I_{\rm outer, \max}$. It initially aligns with the stellar spin and realigns with the outer planet as $J_2$ decreases, which ultimately decreases the mutual inclination  of fixed points to $0^{\circ}$ . 
 
When the $J_2$ precession frequency ($J_2 \sim 10^{-6}$) equals to the slowest inclination oscillation modal frequency ($\sim 0.15^{\circ} yr^{-1}$), the inner equilibrium inclination increase to about half of $I_{\rm outer, \ max}$. 
The slowest inclination oscillation modal frequency can be obtained by calculating the eigenvalue of the following matrix as discussed in \cite{murray_dermott_2000},

\begin{equation}
    \begin{split}
        B_{jj} & = -n_j \frac{1}{4} \frac{m_k}{m_c + m_j} \alpha_{12} \Bar{\alpha}_{12} b_{3/2}^{(1)} (\alpha_{12}), \\
        B_{jk} &= -B_{jj},
    \end{split}  
\end{equation}
 where $n_j$ is the mean motion, $\alpha_{12}$ is the ratio of the semi-major axis (inner to outer planet), $b_{3/2}^{(1)} (\alpha_{12})$ is the Laplace coefficient, and $\Bar{\alpha}_{12} = \alpha_{12}$ or $1$, when $j=1$ or $j=2$.
Note that different combinations of inclinations (i.e., different $\overline{\mathcal{J}_z}$) may have quantitatively different evolution of the fixed point with respect to $J_2$, but they are qualitatively the same.

Figure \ref{fig:fix_point} (b) illustrates the evolution anti-aligned fixed point. The maximum inner planet inclination is used as reference, 
\begin{equation}\label{eq:fix_point_eq}
    I_{\rm inner, max} = \cos^{-1} \left(\frac{\overline{\mathcal{J}_z} -1}{\beta} \right).
\end{equation}
 
It shows that if the outer planet starts aligned with the stellar spin and the inner planet is misaligned, the mutual inclination can stay nearly the same, insensitive to the decaying $J_2$, though it increases slightly.

Figures \ref{fig:fix_point} (a) and (b) can illustrate the inclination evolution when the system is close to the fixed points. However, when the system is ``further" from these fixed points, e.g., the co-planar case, the evolution of inclinations would be somewhat different. Co-planar configuration of the planets could be common assuming in-situ formation or disk migration of the USPs followed by tidal decay \citep[e.g.,][]{Schlaufman_2010, Lee_2017}. Here we run the secular evolution of a co-planar case with initial inclinations $I_{\rm inner,0} = I_{\rm outer,0}=10^{\circ}$ for illustration. A simple $J_2$ model,
\begin{equation}\label{eq:j2_time}
    J_2 = \frac{1}{time} \ yr
\end{equation}
is applied and the system is integrated from $10^4$ to $10^7$ yr. The equation of motion utilizes the Lagrangian planetary equations as shown below \citep[e.g.,][]{valtonen_karttunen_2006}, 
\begin{equation}
\begin{split} 
    \dot{e} &= -\frac{\sqrt{1-e^2}}{na^2e}\frac{\partial \mathcal{H}}{\partial \omega} \\ 
    \dot{I} &= - \frac{1}{na^2\sqrt{1-e^2} \sin I} \frac{\partial \mathcal{H}}{\partial \Omega} + \frac{\cos I}{na^2\sqrt{1-e^2} \sin I} \frac{\partial \mathcal{H}}{\partial \omega} \\
    \dot{\omega} &= \frac{\sqrt{1-e^2}}{na^2e}\frac{\partial \mathcal{H}}{\partial e} - \frac{\cos I}{na^2\sqrt{1-e^2}\sin I} \frac{\partial \mathcal{H}}{\partial I} \\
    \dot{\Omega} &= \frac{1}{na^2\sqrt{1-e^2}\sin I} \frac{\partial \mathcal{H}}{\partial I}
\end{split} 
\end{equation}
the eccentricities are fixed to be zero throughout the evolution  in Section 2. We use the stellar equatorial plane as our reference plane for the Hamiltonian.

Figure \ref{fig:fix_point} (c) shows the secular result of inclination evolution for this co-planar case with a $J_2$ decreasing from $10^{-4}$ to $10^{-7}$. The outer inclination stays nearly the same as the angular momentum of the outer planet dominates, while the oscillation amplitude of inner inclination becomes larger as $J_2$ decreases. For the mutual inclination, it is initially excited up to twice obliquity ($\sim 20^{\circ}$), then its amplitude gradually reduces with decaying $J_2$, and ultimately its value reaches the magnitude of the initial obliquity ($\sim 10^{\circ}$).

For this co-planar case, it might be more intuitive to look at the energy contours shown in Figure \ref{fig:contours}.
The black dots in Figure \ref{fig:contours} display the secular result of the coplanar case. These data points are selected based on the corresponding $J_2$ values which are close to $10^{-4}$, $10^{-6}$, and $10^{-7}$. We can see that the system is captured by the region where the energy contour of the inner planet follows the aligned fixed point while that of the outer planet follows the anti-aligned fixed point. 
As $J_2$ decreases, the role of planet-planet interaction becomes more important,  the aligned fixed point of the inner planet gradually moves from the star-aligned to the planet-aligned, increasing the oscillation amplitude of the inner inclination and decreasing that of the mutual inclination.

\subsubsection{System dominated by the inner angular momentum}\label{sec:sys_b}
 \begin{figure}[ht!]
\centerline{\includegraphics[height=3 in]{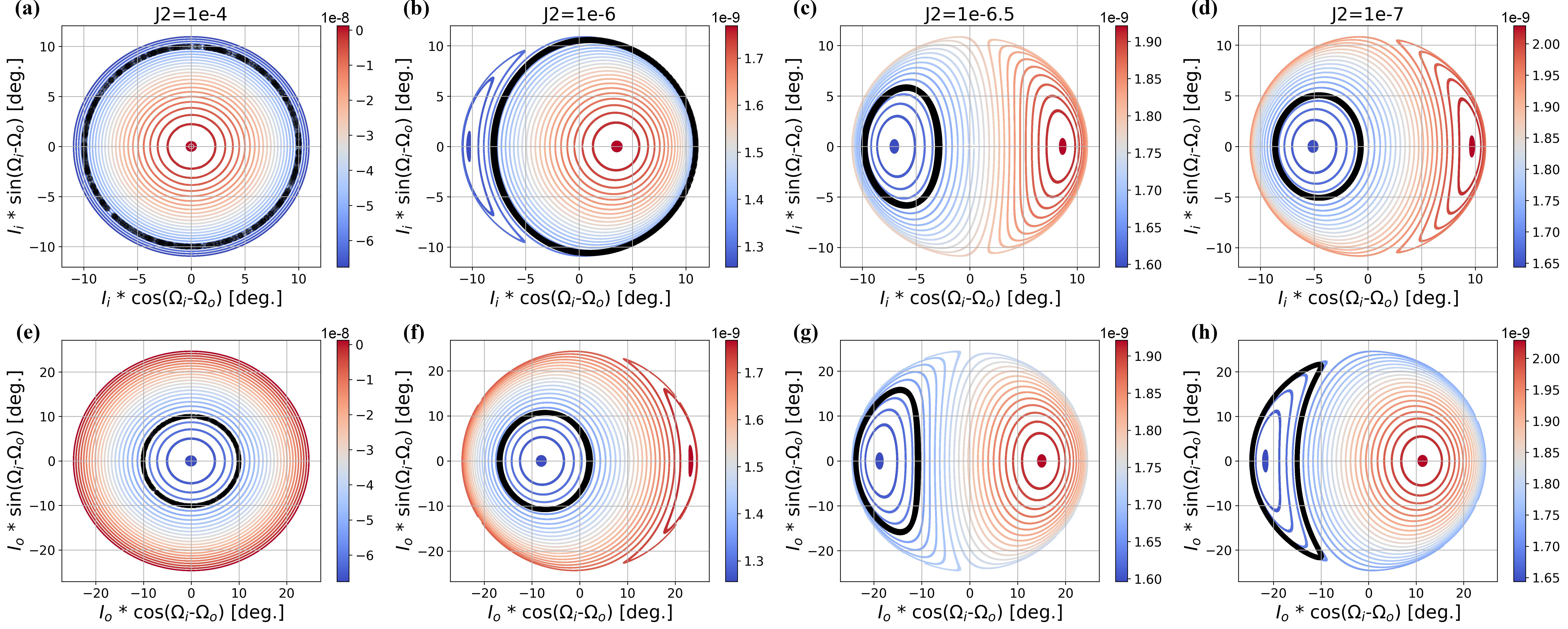}}
\caption{Energy contours of constant Hamiltonian for System B (Equation \ref{eq:H_c}) with different $J_2$ value.  $\overline{\mathcal{J}_z}$ corresponds to $I_{\rm inner} = I_{\rm outer} =10^{\circ}$ and $\beta = 5$. The x axis is $I_{\rm inner} \cos(\Omega_{\rm inner}-\Omega_{\rm outer})$ and the y axis is $I_{\rm inner} \sin(\Omega_{\rm inner}-\Omega_{\rm outer})$ in the first row. For the second row, the x axis is $I_{\rm outer} \cos(\Omega_{\rm inner}-\Omega_{\rm outer})$ and the y axis is $I_{\rm outer} \sin(\Omega_{\rm inner}-\Omega_{\rm outer})$. The color represent the total potential energy $\mathcal{H}_{\rm circular}$ and the black dots show the trajectory of a secular evolution starting with the coplanar configuration ( $I_{\rm inner,0} = 10^{\circ}$ and $I_{\rm outer,0} =10^{\circ}$). \label{fig:contours_large_beta}}
\end{figure}

Next, we show systems where the inner orbital angular momentum dominates (System B, $\beta = 5$). $\overline{\mathcal{J}_z}$ corresponds to $I_{\rm inner}=I_{\rm outer}=10^{\circ}$ and $\beta = 5$. Figure \ref{fig:contours_large_beta} shows the energy contours with respect to different $J_2$.  
In the first row of Figure \ref{fig:contours_large_beta} (for the inner planet), the libration region around $\Omega_{\rm inner}-\Omega_{\rm outer}=180^{\circ}$ appears as an island when $J_2$ is $10^{-6}$. 
As $J_2$ decreases, we can see that the inclination of aligned fixed point  for the inner planet gradually increases while the anti-aligned one decreases. For the outer planet (the second row), the inclination of anti-aligned fixed point increases as $J_2$ decreases, while the inclination of aligned fixed point decreases.

\begin{figure}[ht!]
\centerline{\includegraphics[height=1.8 in]{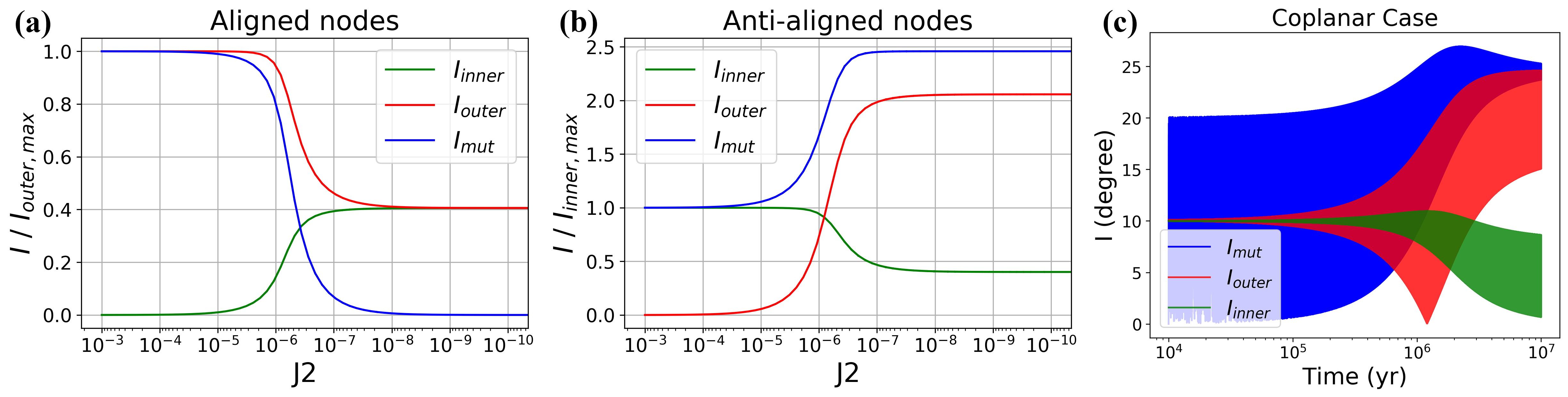}}
\caption{ (a) Evolution of the aligned fixed point for the inclinations based on the Equation \ref{eq:fix_point_eq_aligned} with respect to the decreasing J2. The y-axis is the ratio between the inclination and the initial outer inclination (i.e., when $J_2$ is $\sim 10^{-3}$. (b) Evolution of the anti-aligned fixed point for the inclinations based on the Equation \ref{eq:fix_point_eq_mis} with respect to the decreasing J2. The y-axis is the ratio between the inclination and the initial inner inclination.
(c) Secular evolution for an initially co-planar case with $I_{i,0}=I_{o,0}=10^{\circ}$.\label{fig:fig4_fixed_point}}
\end{figure}

The evolution of fixed points with a decreasing $J_2$ is qualitatively the same as System A which has a much smaller $\beta$, but they are quantitatively different. We show the evolution of the fixed points of System B in Figure \ref{fig:fig4_fixed_point}. The aligned fixed point evolution is displayed in Figure \ref{fig:fig4_fixed_point} (a). It shows that the mutual inclination of fixed points drops to zero as $J_2$ decreases to a small value, which is the same as the case of System A. However, for System B, the changes in the fixed points of inner inclinations are smaller compared to that of the outer inclination.
 This is because the orbital angular momentum of the inner planet dominates in System B, which is the opposite of System A.

The secular result of inclination with an initially co-planar configuration ($I_{\rm inner, 0}=I_{\rm outer, 0}=10^{\circ}$) is shown in Figure \ref{fig:fig4_fixed_point} (c). 
The maximum mutual inclination can be excited to more than $\sim 2.5$ times of the initial obliquity (or the initial average mutual inclination). This corresponds to the secular resonance illustrated in \cite{Spalding_2016}, where the magnitude of final mutual inclination can be excited.
The black dots in Figure \ref{fig:contours_large_beta} display this secular result as trajectories in the plane of energy contours. Initially, the inner planet precesses around the stellar spin when $J_2 = 10^{-4}$. When $J_2$ decreases to $10^{-6.5}$, the inner inclination is captured by the anti-aligned fixed point. The outer inclination keeps following the anti-aligned fixed point which has moved further to the anti-aligned orientation, making the outer inclination increase. As $J_2$ continues decreasing to $10^{-7}$, the anti-aligned fixed point of the inner planet moves toward the origin, leading to the decrease in the inner inclination. Comparing with Figure \ref{fig:fig4_fixed_point} (b) and (c), we can see that the evolution of inclinations after $10^6$ yr in Figure \ref{fig:fig4_fixed_point} (c) nearly matches the evolution of anti-aligned fixed points when $J_2$ is smaller than $\sim 10^{-6}$.

\subsection{Relationship between the final inclinations and $\beta$}\label{sec:beta}

As shown in the previous section, the comparison between System A and B illustrates that different $\beta$ could lead to different inclination evolution. Thus, in the following, we show the relationship between the magnitude of $\beta$ and the final inclinations (i.e., when $J_2$ is small). The evolution can be categorized into three representative regimes, which can be learned from the fixed points. Specifically, the inclination evolution of a system starting with an aligned inner planet and a misaligned outer planet follows the aligned fixed point. The anti-aligned fixed point can tell us the inclination evolution of a system initially with a misaligned inner planet and an aligned outer planet. Finally, for the co-planar configuration, the inclination evolution can be implied by the energy contour as the system could switch the fixed point that it follows. In the following, we discuss the dynamics for these three representative examples as $\beta$ changes.

 \begin{figure}[ht!]
\centerline{\includegraphics[height=1.6 in]{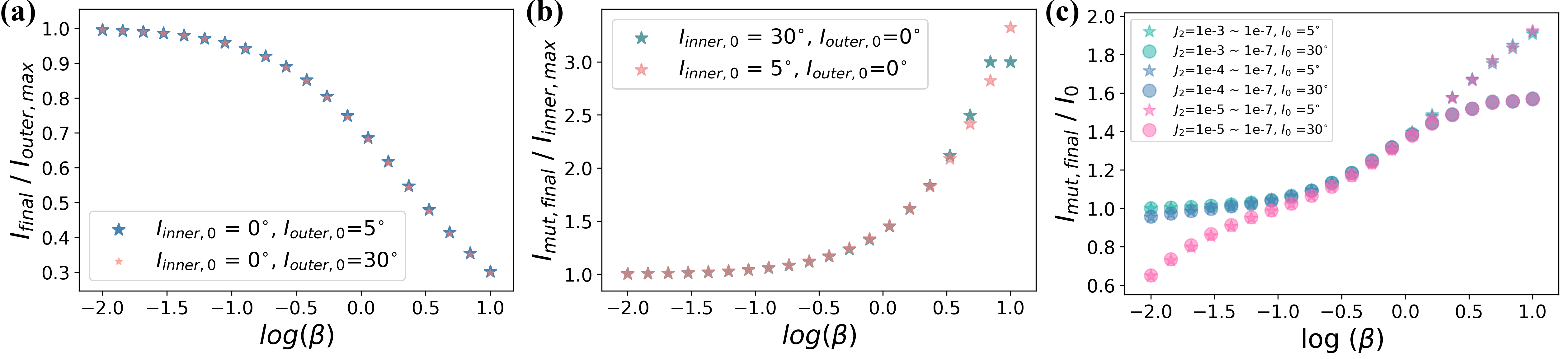}}
\caption{(a) The relationship between the final inner (or outer) inclination and $\beta$. The y axis is the ratio of inclination to the maximum outer inclination (here is $5^{\circ}$ for blue star and $30^{\circ}$ for pink star) for the aligned fixed point. (b) The relationship between the final mutual inclination and $\beta$ for the anti-aligned fixed point. The y axis represents the ratio between the final mutual inclination and the maximum inner inclination.
(c) The average of final mutual inclination with respect to $\beta$ for the initially co-planar case. The y axis is the final average mutual inclination over the initial obliquity. \label{fig:fig5}}
\end{figure}

For the system with the initially aligned inner planet at the aligned fixed point, we can easily read from Equation \ref{eq:fix_point_eq_aligned} that when $J_2 = 0$, then $I_{\rm inner,final} =I_{\rm outer,final}$. Thus the final mutual inclination (the final inclination here refers to the inclination when $J_2 = 0$) is zero regardless of $\beta$. Then, we can estimate the final inclination of the inner and the outer orbits for different $\beta$. Setting two sets of {$I_{\rm inner,0}$, $I_{\rm outer, 0}$} ($I_{\rm inner,0} = 0$ for both sets and $I_{\rm outer,0} = 5^{\circ}, 30^{\circ}$) to initialize $\overline{\mathcal{J}_z}$ , and combining $I_{\rm inner,final}= I_{\rm outer,final}$ and Equation \ref{eq:angular_momentum}, Figure \ref{fig:fig5} (a) shows the ratio between final inner (or outer) inclination and the maximum outer inclination with respect to $\beta$. 
Assuming initially (i.e., when $J_2$ is dominant)  the inner planet is star-aligned, we can see that the ratio decreases as $\beta$ increases regardless of the initial outer inclination.
Increasing $\beta$ means the inner planet possesses more orbital angular momentum. Therefore, when $\beta$ is small (e.g., $\sim 0.01$) the outer inclination nearly stays constant while the inner planet inclination increases and aligns with the outer planet as $J_2$ decreases to zero. When $\beta$ is larger, the outer inclination decreases more (e.g., from the ratio of 1 to $\sim 0.3$ when $\beta = 10$) as the inner planet has more angular momentum than the outer one.

For systems at the anti-aligned fixed point, by setting $J_2=0$ in Equation \ref{eq:fix_point_eq_mis} and combining Equation \ref{eq:angular_momentum}, we can get the final mutual inclination over the maximum inner inclination with respect to $\beta$ as shown in Figure \ref{fig:fig5} (b). For small $\beta$ (0.01 $\sim$ 0.1), the final mutual inclination stays nearly constant with decaying $J_2$.
With a larger initial inner inclination (e.g., $30^{\circ}$ shown in Figure \ref{fig:fig5} (b)), the final mutual inclination does not keep increasing when $\beta \sim 10$. This can be shown mathematically looking at the right hand side of Equation \ref{eq:fix_point_eq_mis}. For the evolution that initially follows the anti-aligned fixed point, the solution can be obtained using the left term ($-\frac{\sin I_{\rm outer}}{\beta} + \sin I_{\rm inner}$) when $J_2$ is dominant. However, when $J_2$ is 0 and the initial inner inclination as well as $\beta$ is large, the solution is determined by the right term ($\sin[2(I_{\rm inner}+I_{\rm outer})]$) and thus independent of $\beta$.

For the co-planar case, the systems are farther from the fixed points, so it is difficult to obtain general results based on the fixed points. Thus, we show the secular evolution results 
to obtain the qualitative trends. Specifically, we set $a_{\rm inner} = 0.01 AU$, $a_{\rm outer}=0.04 AU$, $m_{\rm inner} = 1 M_{\rm Earth}$ with a Sun like star. The mass of the outer planet is adjusted by the value of $\beta$ (see Equation \ref{eq:beta}). We use Equation \ref{eq:j2_time} to evolve $J_2$ here for simplicity, and run simulations with same final $J_2$ ($10^{-7}$) but three different initial $J_2$, $10^{-3}$, $10^{-4}$, and $10^{-5}$. Figure \ref{fig:fig5} (c) shows the averages of final mutual inclination over the initial obliquity ($I_0$) with respect to $\beta$ (``final" here means $J_2$ decreases to $10^{-7}$).
 In general, the final mutual inclination increases as $\beta$ increases, which is consistent with the co-planar results of System A and B shown in Section \ref{sec:sys_a} and \ref{sec:sys_b}. We note that the implied final mutual inclinations could exceed $40^{\circ}$ especially for higher $\beta$ as shown in Figure \ref{fig:fig5} (b) and (c), which could trigger Von Zeiper-Lidov-Kozai oscillations. However, the system with USP tends to possesses a small $\beta$ (e.g., see Figure 6 in \cite{winn_sanchis-ojeda_rappaport_2018}). 
For simplicity, we focus on low inclinations in the following sections with low eccentricity variations. 

\section{Applications to Kepler-653}\label{sec:j2_section}

In this section, we use Kepler-653 as an example to investigate the mutual inclination evolution and constrain its formation mechanism. The parameters of Kepler-653 are obtained from \textit{exoplanets.org} \citep[][]{Han_2014}, which are shown in Table \ref{tab:k653}. 
Kepler-653 has two observed planets, planet c with a mass of $\sim 0.00123 \ M_J$ and a semimajor aixs of $\sim 0.01837 $ AU, planet b with a mass of $\sim 0.0144 \ M_J$ and a semimajor axis of $\sim 0.1183$ AU, and the stellar mass is $\sim 1.02 \ M_{\rm sun}$ with a radius of $\sim 1.19 \ R_{\rm sun}$ \citep[][]{Han_2014, Morton_2016}. The observed mutual inclination is $\sim 12.38^{\circ}$ \citep{Dai_2018}. We first describe the $J_2$ model using \texttt{MESA} simulation \citep{Paxton_2010,Paxton_2013,Paxton_2015,Paxton_2018,Paxton_2019} with Version 20.3.1 \citep{richard_townsend_2020_3706650} in section \ref{sec:j2_model}, then show the secular evolution of the inclinations with three different initial configurations in section \ref{sec:examples}. In section \ref{sec:result}, we introduce an analytical method which can estimate the final mutual inclination 
efficiently and explain how Kepler-653 obtained a large mutual inclination. 

\subsection{J2 Evolution}\label{sec:j2_model}
Focusing on the Kepler-653 system, we obtain a detailed model for the $J_2$ decay. The quadrupole moment decays as the stellar rotation rate decreases due to magnetic braking. The value of $J_2$ can be estimated by (e.g., \cite{Sterne_1939, Spalding_2016})

\begin{equation}\label{eq:J2_eq}
    J_2 = \frac{1}{3} \Big(\frac{\omega}{\omega_b}\Big)^2 k_2,
\end{equation}
where $\omega$ is the stellar angular velocity, $k_2$ is the love number, and $\omega_b$ is the stellar rotational frequency at the break-up. The break-up period is given by 

\begin{equation}\label{eq:J2_eq}
    T_b = \frac{2\pi}{\omega_b} \approx \frac{1}{3} \Big(\frac{M_s}{M_{\odot}}\Big)^{-1/2} \Big(\frac{R_s}{2R_{\odot}}\Big)^{3/2}\  \text{days},
\end{equation}

In this work, the value of apsidal motion constant is obtained by evolving a star with a mass of $1.02\ M_{\rm sun}$ using MESA model \citep{Paxton_2010,Paxton_2013,Paxton_2015,Paxton_2018,Paxton_2019} with Version 20.3.1 \citep{richard_townsend_2020_3706650}, then the Love number $k_2$ is twice of it. We evolve the star from pre-main sequence to 5 Gyr, which is close to the age of the Sun (the age of Kepler-653 is $7.76^{+1.9}_{-2.88}$ Gyr according to \cite{Morton_2016}).
For the stellar rotational frequency $\omega$, it decreases over time mainly due to the magnetic braking, the evolution equation is

\begin{equation}\label{eq:omega_eq}
    \frac{d \omega}{d t} = -\alpha \omega^3,
\end{equation}
where $\alpha = 1.5 \times 10^{-14}$ year, providing a braking timescale of $2 \times 10^{11}$ year for the Sun \citep{barker_ogilvie_2008}. 

\begin{figure}[ht!]
\centerline{\includegraphics[height=2.2 in]{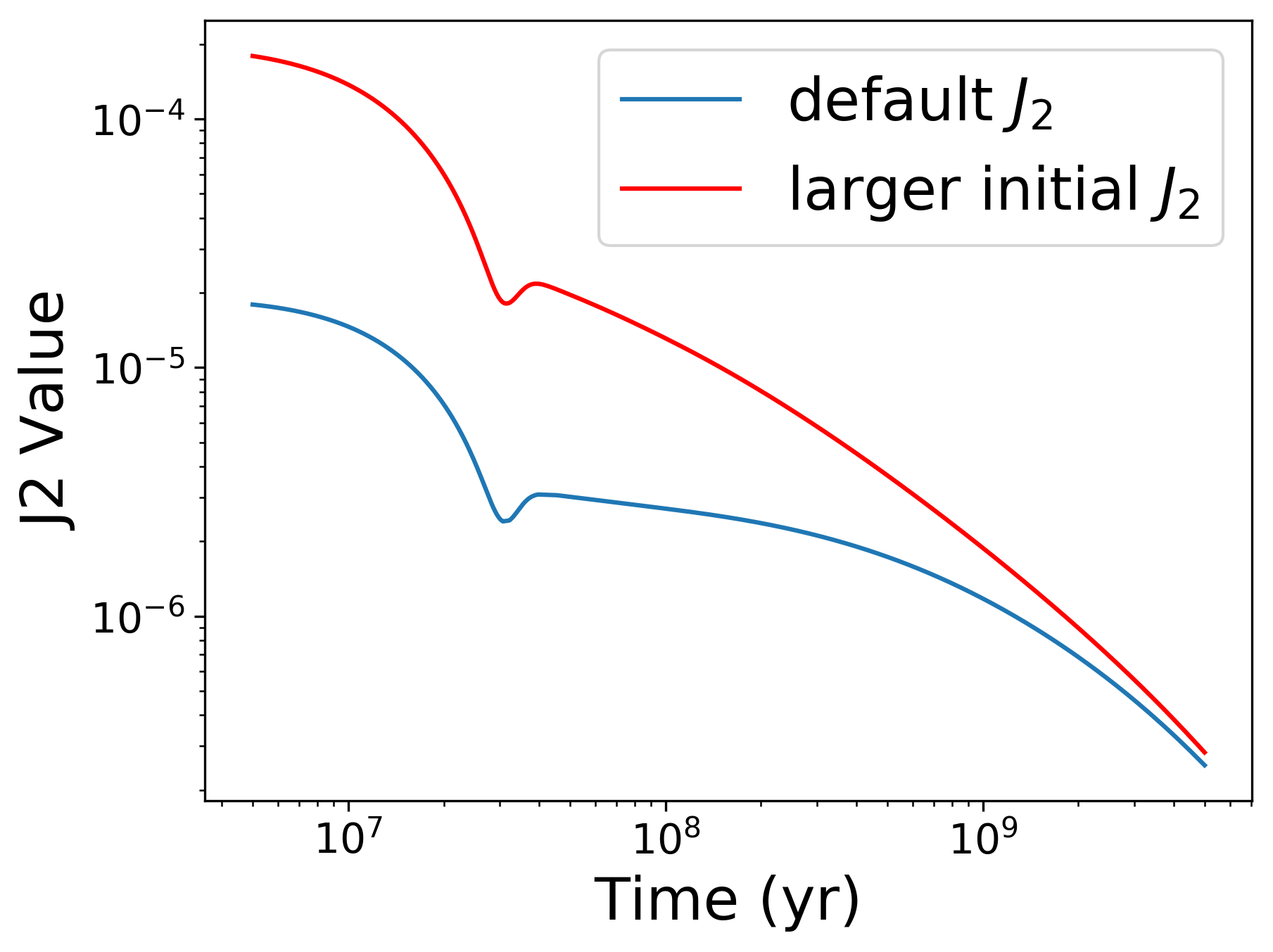}}
\caption{ The evolution of J2 from 5 Myr to 5 Gyr for two different stellar initial rotation periods (10 and 3 days). \label{fig:j2_value}}
\end{figure}

In the following, we consider two situations for $J_2$ evolution.
The first one is the default case, we adopt the rotation period of the Sun, $\sim$ 30 days, at the age of 5 Gyr, and then estimate $\omega$ using Equation \ref{eq:omega_eq} to obtain its rotation period around 5 Myr ($\sim 10$ days). The evolution of $J_2$ from 5 Myr to 5 Gyr is plotted in Figure \ref{fig:j2_value} and color-coded in blue, which decreases from $\sim 1.8 \times 10^{-5}$ to $\sim 2.5 \times 10^{-7}$. There is a ``dip" on the curve of J2 around 30 Myr, which is a time close to the end of pre-main sequence \citep{Iben_1965}.

Because the real rotation period for the star could be faster, e.g., \cite{Matt_2015} shows that the stellar rotation period
with $\sim$ one solar mass in $\sim 2$ Myr Orion Nebula cluster (ONC) ranges from $\sim 1$ to $\sim 10$ days. Thus, we also model $J_2$ with a larger initial value. We let the initial rotation period be $\sim 3$ days and thus the initial $J_2$ is $\sim 10$ times larger than the default one. The evolution is represented in Figure \ref{fig:j2_value} and color-coded in red, which decays from $\sim 1.8 \times 10^{-4}$ to $\sim 2.8 \times 10^{-7}$. We can see that as $J_2$ decreases with time, both $J_2$ reach the same order of $\sim 10^{-7}$.

\begin{deluxetable*}{ccc}
\tablecaption{Parameters of Kepler-653 System \citep{Han_2014, Morton_2016}\label{tab:k653}}
\tablewidth{0pt}
\tablehead{
\colhead{} & \colhead{\textbf{Planet c}} & \colhead{\textbf{Planet b}}
}
\startdata
Mass (\textit{$M_J$}) & $0.00123 \pm 0.00057$ & $0.0144 \pm 0.00197$\\
Semimajor axis \textit{a} (AU) & $0.01837 \pm 0.00031$ & $0.1183 \pm 0.00197$\\
Orbital period (day) & $0.9003765 \pm 3.6 \times 10^{-6}$ & $14.707490 \pm 3.8 \times 10^{-5}$\\
Eccentricity \textit{e} & 0 & 0\\
Argument of Periastron  (\textit{deg.}) & 90 & 90\\
\\
\textbf{Stellar properties}\\
Mass ($M_{\rm sun}$): 1.020 +0.05/-0.04\\
Radius ($R_{\rm sun}$): 1.19 +0.17/-0.16
\enddata
\tablecomments{The minimum mutual inclination of Kepler-653 is 12.38 +1.52/-1.72 degree based on the result of transit light curves fitting \citep{Dai_2018}. }
\end{deluxetable*}

\subsection{Specific Examples}\label{sec:examples}

In this section, we run secular simulations using the default $J_2$ models outlined in the previous section. Specifically, we relax the assumption of circular orbits and run the full secular evolution for Kepler-653 (i.e., following Equation \ref{eq:full_H1}) to study the inclination evolution with different arrivial time (i.e., different initial $J_2$). As shown in Section \ref{sec:sys_a}, with different initial $J_2$ values, the `location' of fixed point varies, changing the distance from the system to the fixed point (this distance is defined as $I_{\rm precess}$). 

 This will result in different final mutual inclinations, and could in turn constrain the arrival time of the planet when compared with observations.

As discussed in section \ref{sec:beta}, systems with the USP planets tend to have small $\beta$, and for Kepler-653, its $\beta$ is $\sim 0.03$. Therefore, we expect that its inclination evolution is similar to that for System A shown in Figure \ref{fig:fix_point}. 
The modal frequency, $\sim -5.9 \times 10^{-3 \circ} yr^{-1}$
matches with the $J_2$ precession frequency when $J_2 \sim 3 \times 10^{-7}$, where the inner inclination of aligned fixed point is expected to increase up to about half of the maximum outer inclination. 

Here we consider three initial configurations, which are motivated by the analysis of fixed point evolution (as shown in section \ref{sec:hami}) and the formation mechanism of USP: 1) an aligned (with respect to the stellar spin) inner planet with a misaligned outer planet ($I_{\rm inner, 0} = 1^{\circ}, I_{\rm outer, 0} = 22^{\circ}$), 2) the co-planar case ($I_{\rm inner, 0}=I_{\rm outer, 0}=7^{\circ}$), and 3) an aligned outer planet with an misaligned inner planet ($I_{\rm inner, 0} = 15^{\circ}, I_{\rm outer, 0} = 1^{\circ}$). 

The co-planar configuration can be formed by the disk migration and tidal decay \citep{Schlaufman_2010} as well as in-situ formation and tidal decay \citep{Lee_2017}. The initial configuration with an aligned inner planet and a misaligned outer planet could be due to 
planet-planet interactions \citep[e.g.,][]{faridani2021hiding}, which could be driven a stellar flyby \citep[e.g.,][]{li_adams_2015}, as well as obliquity tide \citep{Millholland_Spalding_2020}. Moreover, dynamical migrations could lead to the configuration of an aligned outer planet with a misaligned inner planet \citep[][]{Petrovich_2019,Pu_2019}. Finally, the migration during episodic accretion \citep[][]{becker2021migrating} can also contribute to the formation of USPs and the migration timescale is much shorter than the other mechanisms mentioned above.

We set the initial eccentricities of the planets to be nearly circular for all the cases, i.e., $e_{\rm inner, 0} = e_{\rm outer, 0} = 0.01$, and the nodes are initially aligned. We consider two different arrival times (i.e., different values of the initial $J_2$), early arrival (10 Myr, $J_{2} \approx 1.5 \times 10^{-5}$) and late arrival (1 Gyr, $J_{2} \approx 1.2 \times 10^{-6}$). To save the computational time and to ensure an adiabatic change of $J_2$, we scale the planetary system simulation timescale by a factor of five. 

\begin{figure}[ht!]
\centerline{\includegraphics[height=6 in]{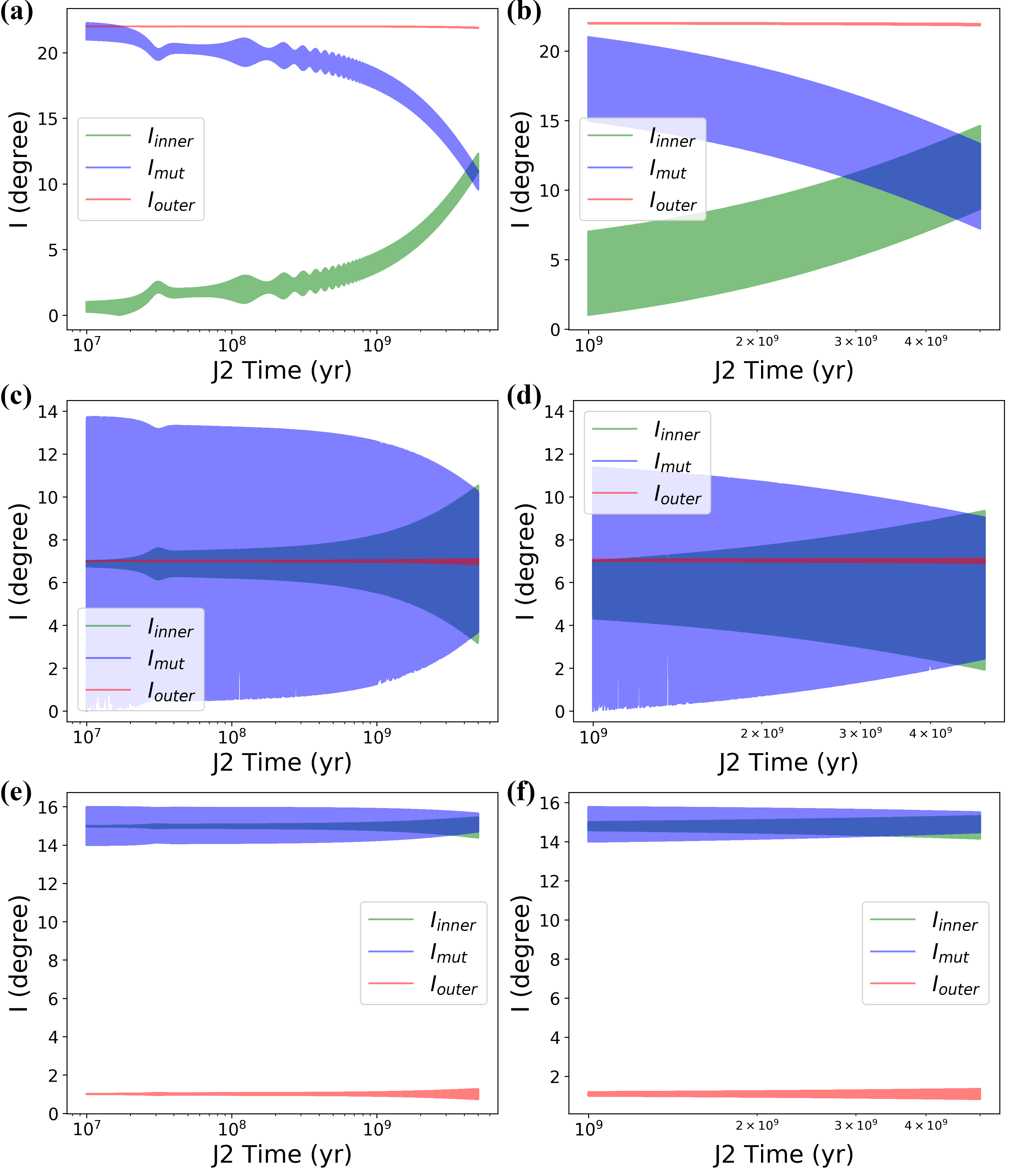}}
\caption{ Specific examples of Kepler-653. First column shows the cases with an arrival time of $10^7$ yr ($J_2 \sim 1.5 \times 10^{-5}$) while second column with $10^9$ ($J_2 \sim 1.2 \times 10^{-6}$) yr. Different rows display different initial configurations. The first row is the case with an aligned inner planet and a misaligned outer planet, $I_{\rm inner,0}=1^{\circ}, I_{\rm outer,0}=22^{\circ}$. The second rows shows the co-planar case with an initial obliquity, $7^{\circ}$. The last row shows the case with a misaligned inner planet and an aligned outer planet, $I_{\rm inner,0}=15^{\circ}, I_{\rm outer,0}=1^{\circ}$. \label{fig:examples}}
\end{figure}

Figures \ref{fig:examples} (a) and (b) display the evolution for the aligned inner planet configuration. Both early and late arrival situations follow the aligned fixed point evolution, i.e., the mutual inclination decreases and the inner planet gradually aligns with the outer one as $J_2$ decreases, which is similar to that shown in Figure \ref{fig:fix_point} (a). The mean inner inclination increases to about half of the outer inclination ($\sim 11^{\circ}$) at 5 Gyr, as the $J_2$ precession frequency is close to the slowest inclination oscillation modal frequency. However, different arrival times lead to different oscillation amplitudes of the inclination. If the planet arrives early around 10 Myr, $J_2$ is still dominant and the system is close to the aligned fixed point. Therefore, the oscillation amplitudes of inclinations are small, as the inner planet precesses around the fixed point with a low $I_{\rm precess}$. When the arrival time is late around 1 Gyr, $J_2$ has decreased and the planet-planet interaction starts to play a more important role in the evolution. The axis that inner planet initially precesses around (i.e., the aligned fixed point) is not aligned with the stellar spin. Thus, the system is captured farther from the aligned fixed point, resulting in a larger oscillation amplitude. Both cases also imply that if the system with such initial configuration arrives its current orbital distance early around 10 Myr, then we cannot observe a minimum mutual inclination of $\sim 12.38^{\circ}$ \citep{Dai_2018}. However, if it arrives late around 1 Gyr, there is a possibility that the mutual inclination of the system agrees with the observation due to the larger oscillation amplitudes. 

Figures \ref{fig:examples} (c) and (d) show us the results for the co-planar case. As $J_2$ decreases, the oscillation amplitude of mutual inclination decreases while that of the inner inclination increases for both the early and the late arrival. However, the final mutual inclination of the early arrival is slightly larger than that of the late arrival. If the planet arrives early around 10 Myr, $J_2$ is dominant and the inner planet initially precesses around the stellar spin with its initial obliquity ($I_{\rm precess} \sim 7^{\circ}$).  For the late arrival, the fixed point of the inner inclination has moved to the orientation of the aligned node (as shown in Figure \ref{fig:contours}). Thus $I_{\rm precess}$ for the late arrival is smaller, which results in the decrease of the amplitude of the final mutual inclination.
Both cases with such a co-planar configuration cannot give us a final mutual inclination that agrees with the observed value, as they both decrease to a magnitude below the observed minimum. This implies that larger initial stellar obliquity is needed for the co-planar case in order to match observation.

For the last configuration (an aligned outer planet with a misaligned inner planet), Figures \ref{fig:examples} (e) and (f) agree with the evolution of the anti-aligned fixed point shown in Figure \ref{fig:fix_point} (b). Following these anti-aligned fixed points, inclinations possess a small oscillation amplitude and stay nearly the same with a decaying $J_2$ as well as different arrival time. They maintain a high mutual inclination, agreeing with the observed value throughout the evolution. Therefore, with this initial configuration, it is always possible for the mutual inclination to match the observation regardless of the arrival time, as long as the initial mutual inclination is larger than the observed value.

\subsection{Secular Parameter Space Exploration}\label{sec:result}
In the previous section, we show the full secular evolution with three different initial configurations: an aligned inner planet with a misaligned outer planet, the co-planar case, and a misaligned inner planet with an aligned outer planet. 
We find that the final mutual inclination decreases for the first two cases, while the third case is not affected by the reduced $J_2$, consistent with our results on the fixed points (section \ref{sec:hami}). Therefore, in this section, we will mainly focus on the first two cases to explore the parameter space of the initial conditions that could agree with the large observed mutual inclination.

We introduce an approximate analytical method to derive the final mutual inclination based on the finding from previous sections, i.e., the evolution of energy contours with decaying $J_2$,  with the assumption of circular orbits. It can efficiently show the final mutual inclination range as well as the likelihood that the mutual inclination agrees with the observed minimum. The steps of this method are detailed below.

1. We can obtain the initial and final $J_2$ value using the assumed arrival time and the final time (e.g., 5 Gyr) with the $J_2$ model, and thus we can find the fixed points ($I_{\rm {inner}, {fixed}_0}$ for the inner inclination) at the initial time as well as that ($I_{\rm {inner}, {fixed}_f}$) at the final time (Equation \ref{eq:fix_point_eq_aligned}). 

2. Then, we can calculate the angle $I_{\rm precess}$ between the fixed point and the initial position of the system using $I_{\rm {inner}, {fixed}_0}$ with the given initial inner orbital inclination and the initial longitude of ascending node. 

3. The area surrounding by the contour is conserved. As an approximation, we assume the contours are circular and $I_{\rm precess}$ are assumed to be constant throughout the evolution as $J_2$ decreases in the adiabatic limit. Then we can estimate the final range of the inner inclination assuming that the azimuthal angle ($\Omega_{\rm azimuthal}$) with respect to the final fixed point is uniformly distributed between $[0^{\circ}, 360^{\circ}]$. 

\begin{figure}[ht!]
\centerline{\includegraphics[height= 3.2 in]{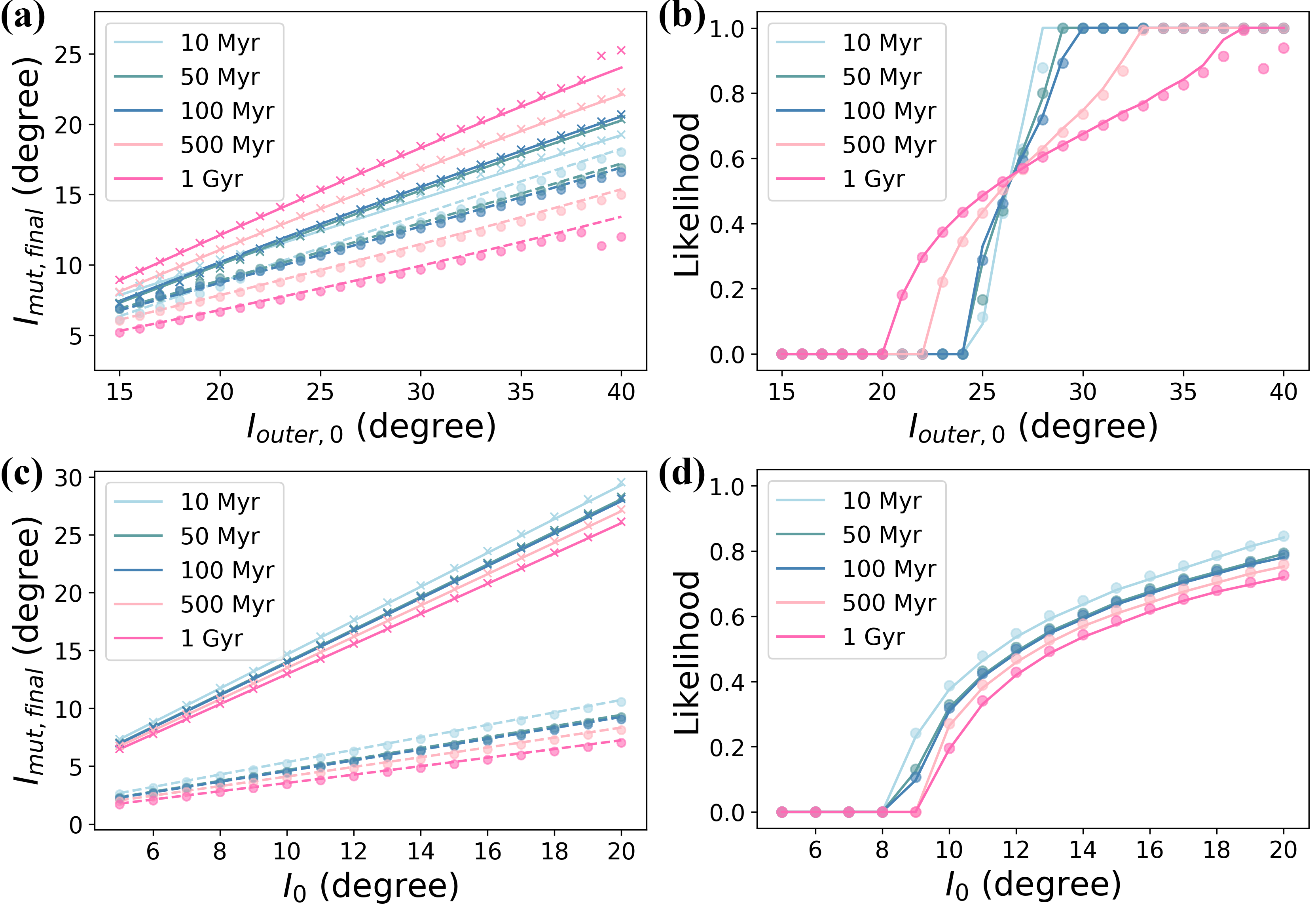}}
\caption{The first and second rows show the results for the aligned inner planet and the co-planar cases, respectively. The first column, (a) and (c), shows the comparison of the final mutual inclination range between the results from full secular evolution and from the analytical method. Different colors show the different arrival times. The solid and dashed lines represent the maximum and minimum values of the final mutual inclination based on the analytical method. The crosses and dots display the maximum and minimum final mutual inclination calculated from the full secular evolution.
The second column, (b) and (d), shows the relationship between the initial outer inclination/obliquity and the likelihood that agrees with observation. The solid line is the result of the analytical method, while the dots show the results from the full secular evolution. 
\label{fig:likelihood_slow}}
\end{figure}

4. Then, the range of the final outer orbital orientation
can be obtained using conservation of angular momentum (Equation \ref{eq:angular_momentum}).  As we know the $I_{\rm {inner}, {fixed}_f}$, $I_{\rm precess}$, $I_{\rm inner}$, and $\Omega_{\rm azimuthal}$, the difference of longitude of ascending nodes (i.e., $\Omega_{\rm inner} - \Omega_{\rm outer}$) can be calculated. Then the final mutual inclination is obtained with the Equation \ref{eq:muti}.
\begin{equation}\label{eq:muti}
    I_{\rm mut} = \arccos\left[\cos{I_{\rm inner}} \cos{I_{\rm outer}} + \sin{I_{\rm inner}} \sin{I_{\rm outer}} \cos(\Omega_{\rm inner} - \Omega_{\rm outer})\right]
\end{equation}
In Figure \ref{fig:likelihood_slow}, we investigate the final mutual inclination using both the analytical method and the secular simulations.

For the secular results, we select the mutual inclination values after 4.9 Gyr as the final range (i.e., minimum and maximum values).
We use the default $J_2$ models described in Section \ref{sec:j2_model}, and consider both the inner aligned as well as the co-planar configurations. For the former, the initial inclination of the inner planet is fixed to be $1^{\circ}$ and the nodes of the two planets are initially aligned ($|\Omega_{\rm inner,0}-\Omega_{\rm outer, 0}|=0^{\circ}$) for all runs. We uniformly choose 26 values from the range $[15^{\circ}, 40^{\circ}]$ for the initial outer inclination. For the co-planar configuration, we choose 16 initial obliquities uniformly between $[5^{\circ}, 20^{\circ}]$. We set the upper bound of the initial outer inclination/obliquity in order to avoid exciting the eccentricity, which could trigger orbital instability\citep[e.g.,][]{Spalding_2016}. The initial eccentricities are $e_{\rm inner} = 0.01$ and $e_{\rm outer}=0.01$ for all runs of the secular simulation.

The first column of Figure \ref{fig:likelihood_slow} shows the maximum and minimum final mutual inclinations as a function of the initial outer planet inclination (the first row) or the initial obliquity (the second row). The first row corresponds to the inner aligned configuration and the second row corresponds to the coplanar case. Different colors represent different arrival time of the planet (corresponding to different initial $J_2$ moments). The solid and dashed lines represent the results of the analytical results, and crosses and dots represent that of the full secular evolution. We can see that our analytical estimations agree well with the full secular evolution results. 

For the initially aligned inner planet case shown in Figure \ref{fig:likelihood_slow} (a), later arrival time gives a larger oscillation amplitude of the final mutual inclination. This is because $I_{\rm precess}$ is larger with a later arrival time as the fixed point evolves further to the right (nodal-aligned direction). Final mutual inclinations of all inner-aligned runs are smaller than their initial values (roughly the value of the initial outer inclination), due to the decrease in $J_2$. Large initial outer inclination and late arrival are needed in order to agree with the observation (e.g., $\gtrsim 20^\circ$, if arrives at 1 Gyr). Compared to the analytical results, secular results show higher oscillation amplitudes with large outer inclinations ($39^{\circ}$ and $40^{\circ}$) and 1 Gyr arrival time. This is likely due to the eccentricity excitation by the secular resonance. 

For the co-planar case displayed in Figure \ref{fig:likelihood_slow} (c), the final mutual inclination is less sensitive to the arrival time of the planet, as the mutual inclination decreases only slightly with a later arrival time. Larger initial obliquity leads to larger final mutual inclination. Note that, if the final $J_2$ is small enough ($\sim 10^{-9}$, close to 0), the final oscillation amplitude of the mutual inclination should be $\sim 0$ \citep[e.g.,][]{Schultz21} for both initial configurations regardless earlier or later arrival time. 

The right column of Figure \ref{fig:likelihood_slow} (b and d) shows the likelihood that the final mutual inclination can match the observed value (i.e., $\geq 12.38^{\circ}$ \citep{Dai_2018}). The dots show the results from the full secular evolution, and are calculated by the percentage of data points that are equal or greater than the observed minimum mutual inclination after 4.9 Gyr. The solid lines represent the results by the analytical method. 

For the inner aligned case, Figure \ref{fig:likelihood_slow} (b) shows that the likelihood generally increases as the initial outer inclination increases, consistent with Figure \ref{fig:likelihood_slow} (a), and a larger initial outer inclination is needed for an earlier arrival time to have a non-zero likelihood.  When the initial outer inclination is smaller than $\sim 26^{\circ}$, later arrival times have higher likelihoods. As discussed in Section \ref{sec:examples} and shown in Figure \ref{fig:likelihood_slow} (a), a later arrival time results in a larger inclination oscillation amplitude, which raises the likelihood. However, when the initial outer inclination is larger ($\gtrsim 26^{\circ}$), a later arrival time tends to have a lower likelihood, which is also due to its larger oscillation amplitude, allowing the mutual inclination to be smaller than the observed minimum.

Figure \ref{fig:likelihood_slow} (d) shows the likelihood for the co-planar case. Similar to the inner aligned case, the likelihood increases with a larger initial obliquity. However, it is 
less sensitive to the arrival time, which is also displayed in Figure \ref{fig:likelihood_slow} (c). A later arrival time requires a slightly larger initial obliquity to have a non-zero likelihood and an earlier arrival time provides a slightly higher likelihood. This is because an earlier arrival time (larger initial $J_2$) results in a larger mutual inclination (as shown in Figure \ref{fig:likelihood_slow} c).

\begin{figure}[ht!]
\centerline{\includegraphics[height= 5.6 in]{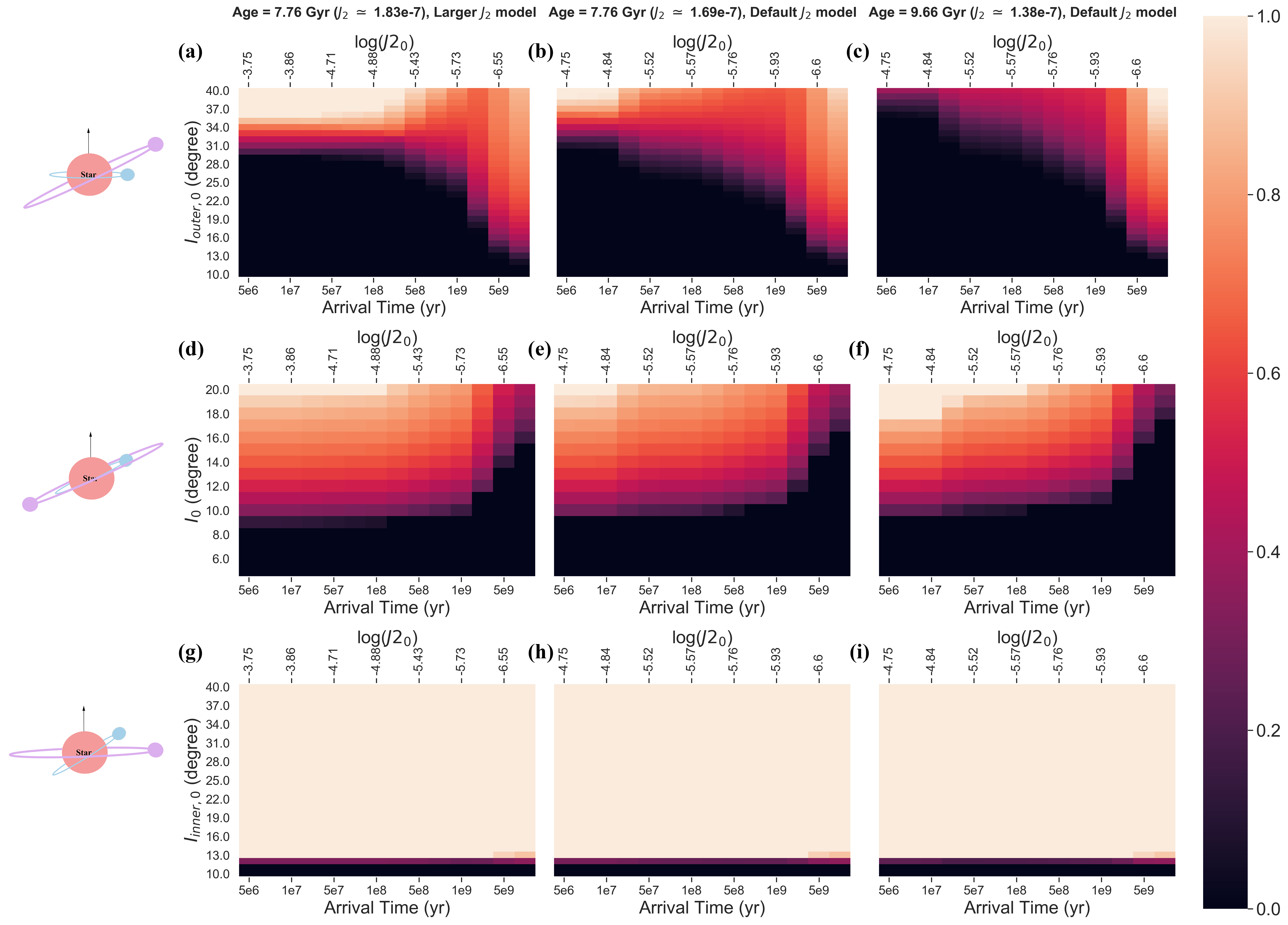}}
\caption{The likelihood of observing a large mutual inclination that agrees with the observation. The color represents the likelihood. The first row corresponds to the configuration of an inner aligned planet with a misaligned outer planet, the second row corresponds to the co-planar configuration, and the third row corresponds to the configuration of a misaligned inner planet with an aligned outer planet. Different columns correspond to different $J_2$ models or stellar ages as shown on the top of each column. The lower x-axis shows the arrival time of the planet and the upper one displays the corresponding initial $J_2$. The y-axis shows the initial inclinations. The initial inner and outer inclinations are both $1^{\circ}$ for the inner aligned and outer aligned cases.
\label{fig:prob_map}}
\end{figure}

Next, we consider a larger parameter space for the initial conditions and estimate the likelihood for the final mutual inclination to agree with observation. We include different planet arrival times, which correspond to different initial $J_2$ moments. This can be used to constrain the formation scenario of the USPs. Different from Figure \ref{fig:likelihood_slow}, where we only considered initially aligned nodes ($\Omega_i-\Omega_o =0$), we study the more general case in the following, with initial node difference uniformly distributed between $0^{\circ}$ and $360^{\circ}$. In addition, given that our analytical method agrees well with the full secular simulation, we use the analytical method to calculate the likelihood for simplicity. We also include the outer planet-aligned case. Note that for the outer aligned case (which follows the anti-aligned fixed point), we use Equation \ref{eq:fix_point_eq_mis} and calculate $I_{\rm precess}$ for the outer inclination with the analytical method. The resulting likelihood maps are shown in Figure \ref{fig:prob_map}. 

We use both larger (first column) and default $J_2$ (middle column) models to see how the different $J_2$ models affect our results. Given that the stellar age of Kepler-653 is 7.76 +1.9/-2.88 Gyr \citep{Morton_2016}, we set the stellar age to be 7.76 Gyr in the first two columns and also consider a different stellar age at 9.66 Gyr in the third column to investigate the effects of the stellar age. For the $J_2$ model, we keep the value of $k_2$ constant after 5 Gyr. The final $J_2$ values for the three columns (from left to the right) are $\sim 1.83 \times 10^{-7}$, $\sim 1.69 \times 10^{-7}$, and $\sim 1.38 \times 10^{-7}$, respectively. 
Different rows show different initial configurations. The first row shows the results with an aligned inner planet and a misaligned outer planet, the second row displays the results for the initially co-planar configuration, and the third row shows the results for the misaligned inner planet with an aligned outer planet. The x-axis shows the arrival time and the upper label presents the corresponding initial $J_2$. For the first row, the y-axis is the initial outer inclination (the initial inner inclination is $1^{\circ}$); for the second row, the y axis is the initial obliquity; and for the third row, the y axis is the initial inner inclination (the initial outer inclination is $1^{\circ}$). Each cell represents the color-coded likelihood.

As shown in the first row, when the inner planet is initially aligned with the stellar spin-axis, it is more likely for the USP to be formed at a later time. If the USP was formed at an earlier time, a large outer inclination is needed (e.g., $\gtrsim 35^{\circ}$ if arrives around 10 Myr when the stellar age is younger or when the star evolves with a faster initial rotation). Specifically, with the larger $J_2$ model (Figure \ref{fig:prob_map} (a)),  the region with the likelihood of unity occurs when the initial outer inclination is larger than $\sim 35^{\circ}$ and the arrival time is earlier than $\sim 250$ Myr. If we consider the default $J_2$ model (Figure \ref{fig:prob_map} (b)), this region becomes smaller and requires an earlier arrival. If the stellar age is older (Figure \ref{fig:prob_map} (c)), the likelihood becomes lower. The likelihood of unity only appears when the arrival time is later ($\sim$ 7.5 Gyr). Despite the differences mentioned above, the boundaries between the impossible region (zero likelihood) and the possible region (non-zero likelihood) show the earlier arrival time needs a larger outer inclination. This is because the later arrival time leads to larger oscillation amplitudes in the mutual inclination, which increases the likelihood to agree with observation. 

For the co-planar case (second row of Figure \ref{fig:prob_map}), higher likelihood appears when the obliquity is larger, and the map is less sensitive with different $J_2$ models and stellar age compared with the inner-aligned case. If the planet arrives before 1 Gyr, the minimum initial obliquities are around $10^\circ$ to agree with observation. Higher initial obliquities are needed to match the observation if the planet arrives later.

For the outer aligned case (the last row of Figure \ref{fig:prob_map}), the likelihoods are nearly unity when the initial inner inclination is larger than $\sim 13^{\circ}$ regardless of arrival times, $J_2$ models, and stellar age. This agrees with the evolution of the anti-aligned fixed point discussed in Section \ref{sec:sys_a}. The variation of $J_2$ does not affect the mutual inclination. Thus, with an aligned outer planet, the final mutual inclination could agree with the observation, as long as the initial mutual inclination is larger than the observed value.

\section{Discussion and conclusion} \label{sec:disc_conclu}
Studies have shown that stellar $J_2$ can enhance the mutual inclination of planetary systems with USPs, and explain the origin of the observed large mutual inclination of USPs with their companions \citep[e.g.,][]{Li_2020,Becker_2020}.
The role of a time varying $J_2$ (as the stellar rotation reduces due to magnetic braking) in the evolution of planetary system has been investigated \citep[e.g.,][]{Spalding_2016, Becker_2020,Schultz21, Brefka21}. Our work sheds the light on how the current observed mutual inclination of the USP system can provide constraints on the initial configuration and formation timing of the USPs. We study the oscillation amplitude of the final mutual inclination under the effect of decaying $J_2$, and investigate the constraints on the USP formation. We found that mutual inclination decreases with the evolving J2 for most of the USP systems (when the inner planetary orbit has lower angular momentum than the outer one). This implies that either (i) the initial obliquities were large, or (ii) the mutual inclination has been acquired late ($\gtrsim 10$) Myrs. 

Specifically, we first analyzed the inclination evolution using the secular approach. Assuming the orbits to be circular, the system could be reduced to one degree of freedom. We find that there are two fixed points: one corresponds to the case where the inner planet is aligned with the stellar spin while the outer planet is misaligned, and the other corresponds to a misaligned inner planet with an aligned outer planet. In general, the mutual inclination of the two planets decreases as $J_2$ decreases following the first fixed point. The mutual inclination evolution around the second fixed point depends on the ratio of the orbital angular momentum of the two planets $\beta=m_i \sqrt{a_i}/(m_o \sqrt{a_o})$. Specifically, for $\beta \lesssim 0.3$, when the inner orbit possesses lower orbital angular momentum, the final mutual inclination stays nearly the same, while for $\beta \gtrsim 1$ the final mutual inclination increases.

Then, we focused on the system Kepler-653 to investigate the evolution of the mutual inclination, including the full secular simulation and relaxing the assumption that the orbits are circular. We note that the results can be slightly different when the mutual inclination becomes large $\sim 40^\circ$, but this does not change our conclusions qualitatively. We included three representative conditions: 1) inner planet aligned with the stellar spin (follows the aligned fixed point); 2) inner and outer planets in the same plane while misaligned with the stellar spin (follows the aligned fixed point but further from it, since $\beta$ is small for Kepler-653); 3) outer planet aligned with the stellar spin (follows the anti-aligned fixed point). Agreeing with the simple secular model with circular orbits and low $\beta$, the mutual inclination decreases as $J_2$ decreases for cases (1) and (2), while the mutual inclination nearly stays the same for case (3).

Specifically, for the inner aligned configuration, USP most likely formed late, in order to reduce the influence of the decaying $J_2$ and enhance the oscillation amplitude of the final mutual inclination. A larger initial mutual inclination ($\gtrsim 35^{\circ}$) is needed if the planet formed early ($\lesssim$ 10-250 Myr depending on the stellar $J_2$ model). The likelihood generally decreases if the initial and final $J_2$ decreases (from Figure \ref{fig:prob_map} a to c). For the co-planar configuration, the USP of Kepler-653 is more likely formed before 1 Gyr with the initial obliquity larger than $\sim 10^{\circ}$. However, if the arrival time is later than $\sim$ 1 Gyr, the required initial obliquity increases. Finally, for the outer aligned configuration, as long as the initial mutual inclination is larger than the observed value, the mutual inclination can always match the observation regardless of the timing of USP formation.

How do the results constrain the formation channels of Kepler-653? If the USP is formed in-situ followed by tidal decay, the initial configuration could likely be co-planar, and the USP likely arrived late as tidal migration could take a long time (order of Gyr) \citep[e.g.,][]{Lee_2017}. Thus, the planet most likely formed with large stellar obliquity e.g., at least $\sim 15^{\circ}$ if arrived around 5 Gyr as shown in Figure \ref{fig:prob_map}.  Note that it is also possible that the innermost planet becomes star-aligned through the adiabatic slow migration, such as the obliquity tide \citep{Millholland_Spalding_2020}. \citet{Millholland_Spalding_2020} showed that the initial obliquity needs to be about $\sim 20 - 40^{\circ}$ to form the USP, which is similar to the required tilts in the top row of Figure \ref{fig:prob_map}, corresponding to the inner-aligned configuration. 
Moreover, if the USP formed during episodic accretion events \citep{becker2021migrating}, the planet tends to arrive early because the USP can migrate in a short timescale with this mechanism. This 
indicates a large initial inclination of the outer planet (e.g., $\sim 34^{\circ}$ if arrived around 10 Myr).  
Additionally, if it is formed via dynamical migration \citep[][]{Petrovich_2019, Pu_2019}, the innermost planet likely becomes misaligned with the stellar spin. Thus, the decrease in $J_2$ does not affect the mutual inclination significantly, and the initial mutual inclination only needs to be greater than the observed value. As a caveat, the disk potential is not included here which could play a role for USPs formed early, and its dispersal timescale is important for the initial configuration set up \citep[][]{Spalding_2020}.
We note that other planetary systems with similar architectural properties can also be analyzed this way to constrain their formation mechanisms, e.g., Kepler-10, and this can also be applied to systems with more than two planets as discussed in \citet{Brefka21}.

\acknowledgments
The authors thank the discussions with Juliette Becker and Josh Winn. This research has made use of the Exoplanet Orbit Database and the Exoplanet Data Explorer at exoplanets.org. G.L. is grateful for the partial support by NASA 80NSSC20K0641 and 80NSSC20K0522. C.P. gratefully acknowledges support from the ANID BASAL projects ACE210002 and FB210003, ANIDMillennium Science Initiative-ICN12009, FONDECYT Regular grant 1210425 and CONICYT+PAI (Convocatoria Nacional subvencion a la instalacion en la Academia convocatoria 2020, PAI77200076). This work used the Hive cluster, which is supported by the National Science Foundation under grant number 1828187.  This research was supported in part through research cyberinfrastrucutre resources and services provided by the Partnership for an Advanced Computing Environment (PACE) at the Georgia Institute of Technology, Atlanta, Georgia, USA.

\bibliography{sample63}{}
\bibliographystyle{aasjournal}



\end{document}